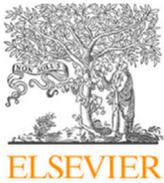
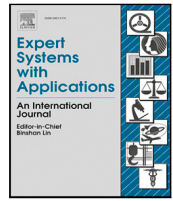

# Simultaneous exercise recognition and evaluation in prescribed routines: Approach to virtual coaches

Sara García-de-Villa [a,1,*], David Casillas-Pérez [b,1], Ana Jiménez-Martín [a,1], Juan Jesús García-Domínguez [a,1]

[a] *Department of Electronics, Polytechnic School, University of Alcala, Alcalá de Henares, 28801 Madrid, Spain*
[b] *Department of Signal Theory and Communications, Rey Juan Carlos University, Fuenlabrada, 28942 Madrid, Spain*

A B S T R A C T

Home-based physical therapies are effective if the prescribed exercises are correctly executed and patients adhere to these routines. This is specially important for older adults who can easily forget the guidelines from therapists. Inertial Measurement Units (IMUs) are commonly used for tracking exercise execution giving information of patients' motion data. In this work, we propose the use of Machine Learning techniques to recognize which exercise is being carried out and to assess if the recognized exercise is properly executed by using data from four IMUs placed on the person limbs. To the best of our knowledge, both tasks have never been addressed together as a unique complex task before. However, their combination is needed for the complete characterization of the performance of physical therapies. We evaluate the performance of six machine learning classifiers in three contexts: recognition and evaluation in a single classifier, recognition of correct exercises, excluding the wrongly performed exercises, and a two-stage approach that first recognizes the exercise and then evaluates it. We apply our proposal to a set of 8 exercises of the upper-and lower-limbs designed for maintaining elderly people health status. To do so, the motion of 30 volunteers were monitored with 4 IMUs. We obtain accuracies of 88.4 % and the 91.4 % in the two initial scenarios. In the third one, the recognition provides an accuracy of 96.2 %, whereas the exercise evaluation varies between 93.6 % and 100.0 %. This work proves the feasibility of IMUs for a complete monitoring of physical therapies in which we can get information of which exercise is being performed and its quality, as a basis for designing virtual coaches.

## 1. Introduction

Nowadays, the number of older adults worldwide is over one billion, which means about the 13.5 % (WHO, 2021) of the population. This aged population leads to a high demand for health care and social services, with the cost that this entails. Health systems are moving from the original disease management to a functional perspective with the objective to get and maintain a healthy ageing (Rodriguez-Mañas, Rodríguez-Artalejo, & Sinclair, 2017). The World Health Organization defines healthy ageing as "the process of developing and maintaining the functional ability that enables well-being in older age" (WHO, 2021).

The regular practice of physical exercise improves the functional capacity of elderly people (Lazarus, Izquierdo, Higginson, & Harridge, 2018) and delays adverse events usually associated with frailty, a clinical syndrome characterized by the vulnerability of the individual and the considerable risk to develop negative health-related events as disability (Rodríguez-Mañas et al., 2013). The combination of strength, resistance, balance and flexibility training in physical routines has shown great improvements in functional capacity (de Asteasu, Martinez-Velilla, Zambom-Ferraresi, Casas-Herrero, & Izquierdo, 2017; Izquierdo, 2019). But these therapies are frequently performed in clinics or hospitals where patients are supervised by therapists or physicians. It compromises patients' autonomy since they have to go expressly to these centers and many times it hardly fits in their activities of daily living. Moreover, due to its high cost, this approach is not suitable for long-term therapies, as those for maintaining elder people health status.

For these reasons, home-based supervision systems, sometimes called virtual coaches, are becoming increasingly important. These virtual coaches contribute to patients' adherence to physical treatments based on exercises (Kyriazakos et al., 2020; Palazzo et al., 2016), which is crucial to obtain the benefits of those long-term therapies. Since they work as a support for the absence of a personal supervisor, virtual coaches have to achieve three main objectives: obtain information about the human motion (Bavan, Surmacz, Beard, Mellon, & Rees, 2019; Mancini et al., 2017); monitor the physical activity, evaluating its performance (Gauthier et al., 2017; Maciejasz, Eschweiler, Gerlach-Hahn, Jansen-Troy, & Leonhardt, 2014; Pereira, Folgado, Cotrim, & Sousa, 2019); and provide feedback to patients (Kyriazakos et al., 2020).

Inescapably, virtual coaches rely on the development of technological solutions to achieve those three objectives. The technological solutions have to be portable and everywhere usable in order to avoid the limitation to controlled environments and increase their availability, so external systems such as the optical ones are dismissed. Portable systems, such as Inertial Measurement Units (IMUs) have increased their use during the last decade because of its low cost and processing simplicity (Lopez-Nava & Angelica, 2016). These sensors measure motions in terms of turn rate and specific force, which is the linear acceleration with the influence of the gravity acceleration. The suitability of IMUs has been demonstrated in several applications related to motion analysis, as joint angle tracking (Fantozzi et al., 2016; Saito, Kizawa, Kobayashi, & Miyawaki, 2020), analysis of kinematic parameters in motor diseases (Goodwin et al., 2021; Romano et al., 2021) and injury risk assessment (Maurer-Grubinger et al., 2021).

Additionally, there are several proposals for the sport motion recognition using IMUs, most of them based on Machine Learning (ML) algorithms. IMUs and ML algorithms have been jointly applied to recognize fitness exercises (Preatoni, Nodari, & Lopomo, 2020), swimming, tennis or basketball (Zhao & Chen, 2020), among others (Cust, Sweeting, Ball, & Robertson, 2019). However, fewer proposals exist for the recognition of motions during the rehabilitation process. As aforementioned, besides the recognition, another problem related with the monitoring of physical routines is the evaluation of the exercises. This evaluation classifies exercises between correct or wrong performances and, in some works, different errors in performance are evaluated. But these works use different ML algorithms for evaluating the performance of particular known exercises individually (Bevilacqua, Huang, Argent, Caulfield, & Kechadi, 2018; Giggins, Sweeney, & Caulfield, 2014; Huang, Giggins, Kechadi, & Caulfield, 2016; García de Villa, Parra, Jiménez Martín, García Domínguez, & Casillas-Perez, 2021; Whelan, O'Reilly, Ward, Delahunt, & Caulfield, 2016). These individual evaluation implies the knowledge of the classifier of the exercise to be evaluated, without requiring its previous recognition.

However, during a remote physical therapy, we do not have such information, and from a practical point of view, not only different exercises are carried out, but also they can be correctly and wrongly performed. Then we have to consider both tasks (recognition and evaluation), that can be carried out separately or as a single task including the whole characterization of the performed exercises. As a consequence, both get more complicated. On the one hand, including correct and wrong performances of the exercises implies that their recognition entails a higher variability than if only accurate performances are taking into account. On the other hand, the evaluation of exercises relies on the correct recognition of motions. An error in the first stage will condition the result of the evaluation. If we combine both stages it results in an increase of the number of data classes, since they do not only include the kind of motions, but also their correctness. In conclusion, the complete characterization of exercises in prescribed routines is a complex task to study. To the best of our knowledge, no previous works deal with the whole exercises characterization as a single classification problem.

In this work, we analyze the problem of exercises characterization, recognizing and evaluating them, with the aim of establishing a first approach for its remote monitoring. We focus on eight upper-and-lower-limb exercises included in a multidisciplinary routine found in the literature (Casas-Herrero et al., 2019), although it can be extrapolated to any exercise routine. The main objective is to characterize these exercises using inertial data from four IMUs placed on the person upper- and lower-limbs, in order to determine which one is being carried out and whether it is correctly or wrongly performed. The inertial data used are publicly available at Zenodo (see Data Availability Section). We evaluate several proposals by using the processed data from IMUs as input data for six different ML algorithms (Bishop, 2006): Support Vector Machines (SVM), Decision Trees (DT), Random Forest (RF), K-Nearest Neighbors (KNN), extreme learning machines (ELM) and Multi-layer Perceptron (MLP). The objective is to know which algorithm performs better.

Our main contribution is the proposal and validation of complete methods that cover the recognition of physical exercises considering also the quality in their performance. For the best of our knowledge, this is the first work that studies the combination of both tasks as a complex one. We evaluate the proposals with different ML algorithms and determine the most suitable one. In this way, we provide insights for the basis of exercises characterization using ML algorithms.

This work is organized as follows: Section 2 gives an overview of the works related to the motions and exercises characterization; Section 3 describes the proposals for the recognition and evaluation of exercises, together with the different ML classifiers considered in this work; Section 4 describes the sensory system, the information about volunteers that participated in this study and the exercises they carried out; Section 5 details the features used as inputs in the proposals, the training and validation setup of the classifiers and the metrics used to compare the proposals; Section 6 details and discusses the results obtained for each proposal; and finally, Section 7 summarizes the main conclusions referred to the exercises recognition and evaluation and the feasibility of performing both tasks by using IMUs.

## 2. Related works

The efficacy of physical physiotherapy programs relies on the patient's adherence and the correct performance of the prescribed routines. Therefore, there is a need for a systematic monitoring of their execution. Recent research has explored the application of technological advances in physical routines monitoring. Video-based and portable technologies are the main alternatives proposed for the monitoring of exercises (Cust et al., 2019).

Video-based solutions are frequently used because of their accuracy and real-time visual feedback. Vision-based methods can be divided into marker-based and marker-less. Marker-based methods provide accurate measurements, about 1 mm, of the position of markers in the 3D space. It is used in Vieira, Sousa, Arsénio, and Jorge (2015) to guide patients during rehabilitation exercises. However, despite their accuracy in motion capture, the difficulties in marker placing and recognition make these systems impractical for virtual coaches (Viglialoro et al., 2019). Furthermore, these technologies have been fused with surface electromyography (sEMG) (Aung, Al-Jumaily, & Anam, 2014) or depth sensors (Colomer, Llorens, Noé, & Alcañiz, 2016) for rehabilitation purposes. However, video technologies are limited to those places where the systems are installed, suffer occlusions and entail patients' privacy concerns (Komukai & Ohmura, 2019; Zihajehzadeh, Member, Park, & Member, 2016). These limitations are overcome by wearable technologies as IMUs.

For a practical characterization of physical routines, recent research has investigated the feasibility of IMUs to provide accurate recognition and evaluation of exercises in different human motion fields, as sports and rehabilitation (Camomilla, Bergamini, Fantozzi, & Vannozzi, 2018; Cust et al., 2019).

Regarding the recognition of the performed exercise or motion, in Zhao and Chen (2020), Zhao and Chen recognized four basketball motions using four IMUs placed on the upper-limbs. They used the mean, variance and absolute value of the maximum fast Fourier transform (FFT) coefficient of each second of the turn rate and specific force signals. They tested combinations of these features and the features obtained with a principal component analysis (PCA) as different possible inputs of SVMs with Gaussian kernel. The results, obtained with four-fold cross-validation, proved that the PCA features provided the highest accuracy, 96 %. Also focused on the upper-limbs, but directly related to rehabilitation, in Bavan et al. (2019), Bavan et al. recognized three shoulder rehabilitation motions performed by patients with subacromial shoulder pain. They used only one IMU on the arm that recorded the turn rate, specific force and magnetic field, which were segmented by selecting unique data segments through a peak analysis function. This study evaluated nine time domain features (mean, root mean square, standard deviation, variance, range, inter-quartile range, percentiles and vector pair Pearson correlation coefficients) and four frequency domain features (maximum frequency component, mean frequency component, energy spectral density, entropy and kurtosis). Four ML algorithms (DT, SVM, KNN and RF) were evaluated using a ten-fold cross-validation and obtained an accuracy over 90 %. However, when they used a leave-one-subject-out (LOSO) cross-validation with the best algorithm (RF), the accuracy decreased to a maximum of 80 %. For the analysis of more complex motions that included the complete body, in Preatoni et al. (2020), they used five IMUs placed on the upper-and lower-limbs and one on the trunk of one side of the body. The motions of study were clean and jerk, box jump, American swing and burpees, and they also took into account and classified the transition intervals when no exercise was being performed. Nine time domain features (mean, standard deviation, root mean square, mean absolute deviation, maximum, minimum, kurtosis, skewness and quartiles) and seven frequency domain features (mean, power, higher frequency, lower frequency, median frequency, mean frequency and spectral entropy) were the inputs for the SVM and KNN algorithms. They evaluated different SVM and KNN kernels and sliding window sizes until 600 ms. They found that the cubic kernel SVM with 600 ms window length obtained the best results for the five-fold cross-validation, with an average accuracy of 99.1 %. With a LOSO cross-validation, this algorithm obtained an average accuracy of 97.6 %.

On the other hand, for the exercises evaluation, the performance of a variety of exercises have been individually assessed. In Whelan et al. (2016), they evaluated the lunge exercise using five IMUs on the lumbar spine and lower-limbs. They used sixteen features per signal (signal peak, valley, range, mean, standard deviation, skewness, kurtosis, signal energy, level crossing rate, variance, first and third quartiles, median and the variance of both the approximate and detailed wavelet coefficients). The binary classification between correct and incorrect performances was evaluated using RF and achieved an accuracy of 90 %. They also analyzed the classification of the specific deviations, as external rotation of foots or short or long starting stances, and found an accuracy around 70 %. Similarly, in Kianifar, Lee, Raina, and Kulic (2017), they studied the single-leg squats exercise evaluation using three IMUs on the low back and on one leg. They used an extended Kalman filter with a biomechanical model proposed in Lin and Kulić (2012) to estimate the human pose. As in Whelan et al. (2016), signals are segmented into exercises repetitions, but in this case only time domain features are used (root mean square, standard deviation, variance, mean, mean absolute deviation, skewness, kurtosis, range, minimum, and maximum). They used a LOSO cross-validation for the binary classification between correct and wrong performances, and they reported an accuracy of 90 % with Naive Bayes (NB), closely followed by SVM, which obtained 89 % of accuracy.

A variety of exercises have been individually evaluated in other studies, based on the knowledge of the type of exercise executed. In Giggins et al. (2014), they used a logistic regression to individually classify between correct and incorrect variations of seven leg exercises. Ten features (mean, standard deviation, skewness, kurtosis, signal energy, level crossing rate, signal range, first and third quartiles and the variance of the wavelet coefficients) were obtained from the measured turn rate, specific force and from the estimated acceleration magnitude and orientation angles, pitch and roll. They monitored the motions with three IMUs placed on the thigh, shin and foot and reported an accuracy between 81 and 83 %. Using the same placement of IMUs on legs, in Huang et al. (2016), more ML algorithms were applied to evaluate a set of seven leg rehabilitation exercises in order to study the optimal IMU placement and combination. In that work, they evaluated logistic regression, together with DT, MLP, SVM, RF and Adaboost classifiers, which combine different ML algorithms to improve their final classifications, and averaged the metrics reported by all the methods. They segmented signals into exercises repetitions and obtained ten time domain features (mean, standard deviation, skewness, kurtosis, maximum, minimum, range, first and third quartiles and cross-correlation), sixteen coefficients of the FFT as frequency domain features and thirty-two wavelet coefficients as time–frequency features. They obtained an averaged accuracy for all the classifiers between 78–97 % in the exercises evaluation. More recently, in Bevilacqua et al. (2018), four different knee rehabilitation exercises were evaluated. To do so, they used a single inertial sensor placed on the shin that measured the turn rate and specific force, and they estimated the turn rate magnitude and the pitch and roll angles. They segmented the exercises repetitions and obtained fifteen time domain features (mean, median, standard deviation, variance, range, kurtosis, skewness, maximum, minimum, positive mean, negative mean, sum of absolute differences, first and third quartiles, and the correlation index between pitch and roll signals) and twenty-five frequency domain features (energy, energy ratio, energy average, harmonic ratio, energy entropy, and the first 20 coefficients of the FFT). They achieved a binary classification using RF and DT with accuracies that ranged between 88–97 %.

IMUs have also been used for the upper-limb exercises evaluation. In Pereira et al. (2019), Pereira et al. combined two inertial sensors with sEMG sensors to supervise two upper-limb exercises and one lower-limb exercise. The inertial sensors were placed on the arm in all exercises whereas the sEMG sensors were placed according to the exercises on the back or lower-limbs. Three statistical features (skewness, kurtosis and histogram) and nine time domain features (mean, median, maximum, minimum, variance, temporal centroid, standard deviation, root mean square, and auto correlation) were used. Feature selection was based on the study of their correlations. They carried out the exercises evaluation with DT, KNN, SVM and RF and obtained an accuracy about 92 % with all the classifiers.

With respect to the gait assessment, in Alcaraz, Moghaddamnia, and Peissig (2017), they studied the quality of gait in order to classify 30 volunteers as healthy or unhealthy. They used seven IMUs placed on the lower-limbs and the lumbar zone to record the turn rate and specific force and fuse their signals to estimate the joint angles through an EKF. They used nine features (motion intensity, peak asymmetry factor, step period, stride period, regularity, sum of power spectral density, spectral entropy, sum signal-to-noise-modulation-ratio and wavelet entropy) for the classification. They applied a Linear Discriminant Analysis (LDA), PCA and NB to obtain accuracies of 100 %, 86 % and 100 %, respectively.

Finally, the authors individually evaluated seven upper-and lower-limbs exercises in García de Villa et al. (2021). Four IMUs recorded the inertial data of the volunteers during the physical routine execution. We used four time domain features (mean, standard deviation, maximum and minimum) of the turn rate and specific force signals as inputs for NB, SVM, RF and KNN. Signals were segmented by a sliding window whose size was adapted to each exercise and the algorithms were evaluated with a random cross-validation with ten iterations. SVM reported the highest accuracy, precision, sensitivity and specificity in the exercises evaluation, with an accuracy between 98–99 %.

As can be inferred, there is a great diversity of algorithms used in the exercise recognition and evaluation. Although, according to Camomilla et al. (2018) and Cust et al. (2019), the most promising algorithms for these objectives are SVM, RF, KNN and neural networks. A variety of features are also extracted from the IMU signals, although the most common are the statistical features in the time domain, such as mean, standard deviation and maximum and minimum of signals. In addition, signal segmentation is approached from two main perspectives: window-based and repetition-based. Repetition-based segmentation involves further signal processing to detect the start and end of repetitions, and window-based segmentation involves determining the most appropriate window size. In the methods found in the literature, window sizes go up to 6 s, although in Banos, Galvez, Damas, Pomares, and Rojas (2014) the window interval of 1–2 s was shown to be the best trade-off solution between accuracy and speed of activity recognition.

This review of the state-of-the-art approaches highlights the need of studying the recognition and evaluation of exercises as a single and complex task, specially for the monitoring of prescribed routines related to health. Virtual coaches, aimed for being used in unsupervised environments, are required to provide a complete characterization of the executed routines. Therefore, it is important to find the most suitable approach for this complex characterization.

## 3. Methods

The main goal of our proposal is to determine which exercise is carried out by a person in a therapeutic session and whether it is being performed according to its prescription (correct performance - C) or not (wrong performance - W). We refer to the process of determining the exercise as *recognition*, whereas the *evaluation* corresponds to the performance assessment, as correct (C) or wrong (W). Thus, our proposals both recognize and evaluate the exercises by measuring the turn rate and acceleration with four IMUs placed on the body of volunteers.

We detail three different approaches in Section 3.1. We relate the formal classification problem with the objective of exercises characterization in Section 3.2 and we provide a brief explanation of the ML algorithms used in this work in Section 3.3.

### 3.1. Proposals for the exercises recognition and evaluation

Simultaneous exercise recognition and evaluation is a complex problem to solve. To deal with this issue, we propose three different approaches with the only *prior* knowledge of the type of exercises included in the dataset. Since the applied ML approaches are highly non-linear, the three proposals have to be characterized because they are not expected to be equivalent. These proposals are explained in the following:

- The first proposal, called "ReEv", makes the recognition and evaluation in one single step providing as outputs the type of exercise and its correctness. Its working scheme is shown in Fig. 1. In this way, with one single classification process we obtain the complete characterization of the performed exercise. However, the number of classes doubles because we have the correct and wrong performance of exercises. That increment of classes complicates the classification task and, as a consequence, is expected to increase the error rates.
- From a practical point of view, we can assume that the recognition of exercises is relevant only if they are correctly performed. Therefore, the characterization of motions, i.e. determining the motion angles or the number of repetitions, is specially interesting in the correct repetitions. On the contrary, the wrongly performed exercises are required to be detected in order to get information about the correct comprehension of the description of exercises. The wrongly performed exercises of lower-or upper-limbs drive to quite similar features, easy to confuse even by humans, what reduce the accuracy rates. Our second approach, called "ReC-W", tries to surpass this issue by eliminating of the recognition process the wrongly performed exercises.

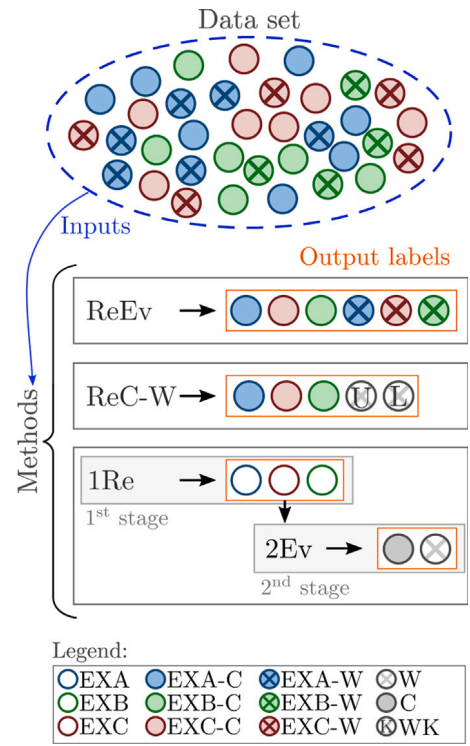

**Fig. 1.** Scheme of the different classification approaches for the recognition and evaluation of exercises used in this study: ReEv, ReC-W and 1Re-2Ev. The scheme shows three different exercises, used as example: exercise A (EXA), exercise B (EXB) and exercise C (EXC); including also their correct (-C) and wrong (-W) performance label. Labels correct (C) and wrong (W) after the recognition of exercises are gray depicted, similar to the wrong label which specifies the kind of limb moved, called WK. In this label, K changes according to the limbs moved during the exercises, being WU when it refers to upper-limb exercises and WL when it does to lower-limb exercises.

The main difference of ReC-W with respect to the first proposal, ReEv, is that ReC-W considers all wrongly performed exercises as two kind of motions, those performed with the upper-limbs and those performed with the lower-limbs. So this method only recognizes the correctly performed exercises, but assigns the generic labels, WU and WL (Wrong Upper-limbs, Wrong Low-limbs), to the wrong performances, as depicted in Fig. 1.
- The last proposal, called "1Re-2Ev", divides the complete process of determining the kind of exercise and its performance quality into two different stages of classification. It is based on the hypothesis that separating the recognition and the evaluation into two different stages, both classifications would improve their accuracy rates. This would be a consequence of three different facts: *(1)* the reduction of the number of classes; *(2)* the increase of variability of each class in the recognition (by the mix of correct and wrong classes); and *(3)* the simplification of the evaluation of each exercise separately after its previous recognition.

Then, we separate both classifications as follows: the initial stage consists in a multi-class classification for the recognition of the exercise; and the second stage evaluates the recognized exercise making a binary classification, as schematized in Fig. 1.

### 3.2. Classification problem

Formally, we consider a set of input–output pairs $\mathcal{D} = \{(\mathbf{x}_i, y_i)\}_{i=1}^{N}$ where $\mathbf{x}_i \in \mathbb{R}^n$ are the $N$ samples of the input feature space obtained from the IMU signals recorded during the exercises and $y_i \in C = \{C_j \mid 1 \leq j \leq J\}$ are the class to which these features correspond. The number of classes $J \in \mathbb{N}$ depend on each proposal. ML classifiers look for a decision function $f$:

$$f : \mathbb{R}^n \to C \qquad (1)$$
$$\mathbf{x} \mapsto y = f(\mathbf{x}, \omega)$$

which given a sample, that in this work contains features from IMU signals, determines the output class, i.e. the one that includes the kind of exercise and its performance correctness.

The so-called parametric ML methods are characterized by a set of parameters $\omega$. During the training process, the ML classifier finds the parameters $\omega$ that best fit the given training data set. The aim of these methods is to find a function $f$ capable of generalizing its good accuracy to the given new data, which corresponds in this study to a person motion features recorded by the IMUs.

### 3.3. ML algorithms evaluated

We evaluate the performance of the following ML algorithms: SVM, RF, KNN, ELM, MLP and DT. All of them are supervised methods which require a labeled data set to be trained. We choose SVM, RF, KNN and the two neural networks, MPL and ELM, because they are the most promising algorithms for the sport monitoring and performance evaluation (Camomilla et al., 2018) and recognition (Cust et al., 2019). Besides, we include DT as baseline method.

SVMs are classifiers which look for the maximum separation among different classes, i.e. their decision function is a separation hyperplane that maximally separates samples from different classes (Scholkopf & Smola, 2018). Usually, SVMs apply appropriate non-linear maps to the input space $\phi : \mathbb{R}^N \to \mathbb{R}^p$ in order to guarantee that the transformed samples are more likely to be linearly separable in a higher-dimension feature space $\mathbb{R}^p$. This is the so-called *kernel trick*, which we employ to improve the performance of the classifier.

DTs are non-parametric methods based on simple decision rules inferred from data features (Breiman, 2001), but their overfitting problem is widely known. RFs are ensemble learning methods that construct $T$ classification decision trees to predict the outputs (Breiman, 2001). They fix the characteristic overfitting problem of individual DT.

RF is one of the most accurate classification algorithms, with good scalable properties: it efficiently deals with large amount of data and multiple input variables without consuming lot of resources such as memory. RFs are trained by the bootstrap aggregating technique (Breiman, 1996), selecting random feature trees during the training process.

KNN classification is a non-parametric ML method which finds a group of $k$ objects in the training set which are the closest to the test object (Shakhnarovich, Darrell, & Indyk, 2008). Frequently, it uses the Euclidean distance, weighting the importance of each feature, which is the distance used in this study. The assignment of a specific class is based on the predominance of a particular class in its neighborhood. The $k$ parameter specifies the size of the neighborhood, which votes for labeling the input data.

MLPs are a kind of feed-forward Artificial Neural Network (ANN) organized as a set of sequentially interconnected layers (Bishop et al., 1995; Kubat, 1999). Each layer is fully connected, which means that all neurons of a layer have links to the neurons in the previous one, through which they receive information, emulating the synaptic links of the human brain. Links have associated weights that adjust the propagation of the information to the output. MLP has a high capacity of generalization, but suffers from overfitting if the number of layers, or neurons in each layer, is not well chosen. During the training process, the different algorithms search the best combination of link weights in order to optimize a goodness-of-fit function. The sequential network topology benefits the efficiency of the optimization methods. Optimization methods such as the backpropagation algorithm combined with the Levenberg–Marquardt algorithm (Levenberg, 1944; Marquardt, 1963) are possible due to this topology. In this work, both methods have been used for training the MLP.

ELM is a special kind of multi-layer perceptron, with one single hidden layer, characterized by being trained with a method computationally faster than the traditional backpropagation method (Huang, Zhou, Ding, & Zhang, 2011; Huang, Zhu, & Siew, 2006). The ELM training process randomly chooses the link weights of the hidden neurons, that frequently follows a uniform probability distribution. The weights of the links that connect the hidden layer to the output are computed establishing a linear least-squares problem which is solved calculating a fast pseudo-inverse, which considerably reduce the computation time during the training. ELMs have less capacity of generalization than MLPs, but suffer less overfitting during the training. The number of neurons of the hidden layer is the only hyperparameter to be determined.

## 4. Experimental protocol

The data for this study were recorded by inertial sensors placed on the volunteers' bodies while they performed the exercises commonly found in physical therapies. In this section, we detail the study population, the sensory system and the studied exercises.

### 4.1. Study population

Thirty volunteers participated in this study, 13 of them were women and the other 17 were men. They were all healthy people aged between 22 and 70 years old. Ten of them were between 20 and 29 years old, and the other forty were groups of five people for each ten years of age. In average, they were $169.1 \pm 7.9$ cm tall and had a weight of $69.5 \pm 10.3$ kg. In this way, we evaluate the proposed methods in a set of volunteers with variability in their age and anthropometric measures. Guadalajara University Hospital approved the study protocol (Institutional Review Board No. 2018.22.PR, protocol version V.1. dated 21/12/2020), and a written informed consent was obtained from all participants.

### 4.2. Sensory system

During the performance of the physical routine, volunteers' motions were recorded with four IMUs. We used the commercial IMUs called *NGIMU*, by *X-io Technology* (x-io Technologies Limited, 2021), which has a size of $56 \times 39 \times 18$ mm, what makes them practical for wearing during the performance of the studied motions. These IMUs include a 3-axis gyroscope, accelerometer and magnetometer, which have a range of $2000\,°/s$, $16$ g and $1300\,\mu T$, respectively. In this particular work, we only use the gyroscope and the accelerometer that have a 16-bit resolution, and a maximum sample rate of 400 Hz. For the experiments, these devices measure the turn rate and specific force in each axis (six signals) during the exercises performance at a sample rate of 100 Hz. The signals of each device are stored into one micro SD card and processed off-line. However, the *NGIMU* has wireless communication what would meet the requirements of virtual coaching. The inertial data recorded during the experiments are publicly available at Zenodo.

The experiments were carried out in the MoCap laboratory of the University of Alcalá. This is a controlled environment where volunteers were recorded one at a time. IMUs were placed on the volunteers' thighs and shins during lower-limb exercises and on arms and forearms during upper-limb exercises. We used Velcro straps to secure the attachment of the sensors, while ensuring that their placement and tightness did not affect their freedom of movement. On the lower-limbs, we placed IMUs in the anterior surface of their limbs and, on the upper-limbs, IMUs were placed on the exterior lateral location, as Fig. 2 shows. In all cases, the orientation of IMUs is the same, with the $X$-axis pointing to the ceiling when volunteers kept standing with their arms along their bodies and their hands pointing to the floor. We selected those sensors location on the body because of the easiness of their placing.

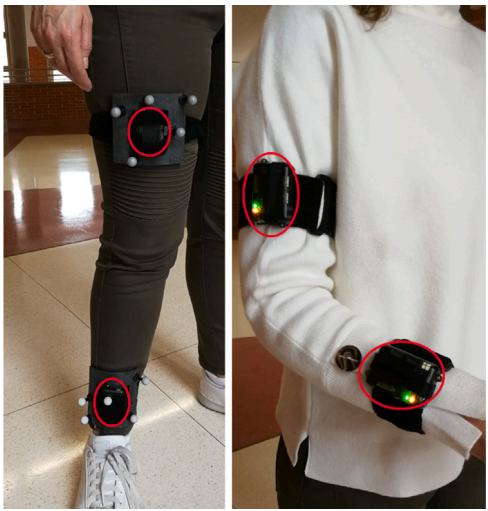

**Fig. 2.** IMUs placed at the right lower-limb and at the right upper-limb, marked with red circles. On the left limbs, there are two more IMUs at similar placements than in the right upper-and lower-limbs, respectively. Passive reflector sensors, which are commonly used to obtain optical reference data in biomechanical studies as in García-de Villa, Jiménez-Martín, and García-Domínguez (2021), are also shown. However, the optical data are not used in this study.

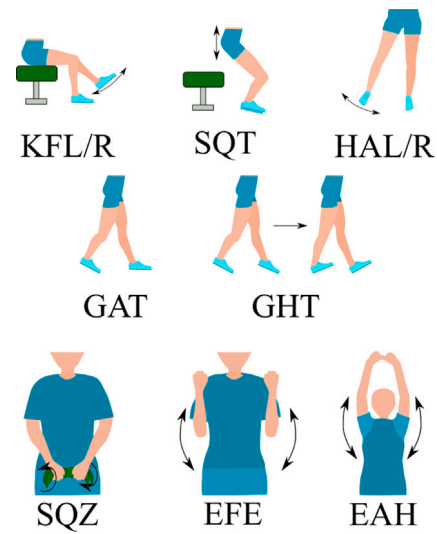

**Fig. 3.** Exercises considered in this study, based on Casas-Herrero et al. (2019). The first row includes knee flex-extension (KFL/R), squats (SQT) and hip abduction (HAL/R). The row below contains natural gait (GAT) and heel-tiptoe gait (GHT). Finally, the column in the right shows squeezing (SQZ), elbow flex-extension (EFE) and extension of arms over head (EAH).

Our proposals rely on maintaining this location of IMUs because the training data only covers this configuration. As a consequence, following applications of these methods are limited to data obtained using the same IMU locations on the body.

*4.3. Experimental tests: exercises*

A set of eight exercises were carried out by the volunteers, which are focused on the lower-or upper-limbs. Since two of them divides into their performance with the corresponding side of the body, right (R) of left (L), we study 10 types of exercises. These exercises consist in repetitive motions, commonly prescribed to older adults, that have to be performed in a specific way. In this study, volunteers mimicked the instructions found in Casas-Herrero et al. (2019) in order to perform the exercises explained in the following. First, they carried out a set of lower-limbs exercises:

- Knee flex-extension (KFL/R): seated on a stable surface, from the initial position of 90° of knee flexion, keeping the left leg still, the right one moved until its extension and returned to knee flexion. After all repetitions moving the right leg, the right leg remained still and the left one moved, as shown in Fig. 3-KFL/R.
- Squats (SQT): from standing position, volunteers made the motion of sitting on a chair and, when touching the chair with their back thighs, they stood up again. This exercise is depicted in Fig. 3-SQT.
- Hip abduction (HAL/R): standing up, keeping the left leg still, the right one moved doing an abduction–abduction with the leg straight, as Fig. 3 schematizes. After all repetitions moving the right leg, the right leg remained still and the left one moved.

Besides these exercises of legs, we studied different variations of gait (see second row of Fig. 3). We consider the gait variations as leg exercises because IMUs were placed on the volunteers' lower-limbs during their execution, even though they are more complex. Volunteers performed two gait variations:

- Gait (GAT): volunteers walked freely in the room.

- Gait with heel-tiptoe (GHT): during walking, volunteers placed first the heel on the floor and then they stood on tiptoe. Keeping their weight on their tiptoe, they placed the other heel on the floor and repeated the motion.

Furthermore, volunteers performed the following exercises of the upper-limbs, which are depicted in the third row of Fig. 3:

- Squeezing (SQZ): using a clothing and keeping arms straight forward, wrists moved anti-symmetrically squeezing the clothing.
- Elbow flex-extension (EFE): both arms moved from the straight position to the maximum flexion of elbows, keeping the shoulders still.
- Extension of arms over head (EAH): with both hands together, arms, as straight as possible, made an arch until reaching the maximum elevation of hands.

Volunteers repeated each motion between 10 and 20 times, according to their age, except for the squats. The squats were repeated only between 7 and 15 times to prevent a wrong performance of the exercises caused by physical fatigue, as observed during the experiments with more repetitions. Volunteers performed four times each exercise series. Two of these series consisted in the corresponding repetitions properly done; and the other two series were wrongly performed, with a total motion freedom to modify the original exercise. The last two series are the ones labeled as wrong. Since volunteers made various motions for the wrong performances, data of this wrong exercises include a high variability. The gait exercises are an exception since we consider only one kind of wrongly gait variation, in which volunteers walked freely mimicking a tired gait or joint locking. So, gait divides into three variations: correct GAT, wrong GAT and GHT.

**5. Data analysis**

Fig. 4 depicts the overview of the data analysis for the exercises recognition and evaluation using signals from four IMUs. The four dotted rectangles refer to the following sections: Section 5.1 details the signal segmentation and feature extraction; Section 5.2 defines the classes for each proposed method; Section 5.3 introduces the optimization parameters or configuration for the ML algorithms; and Section 5.4 describes the metrics used to evaluate the proposals.

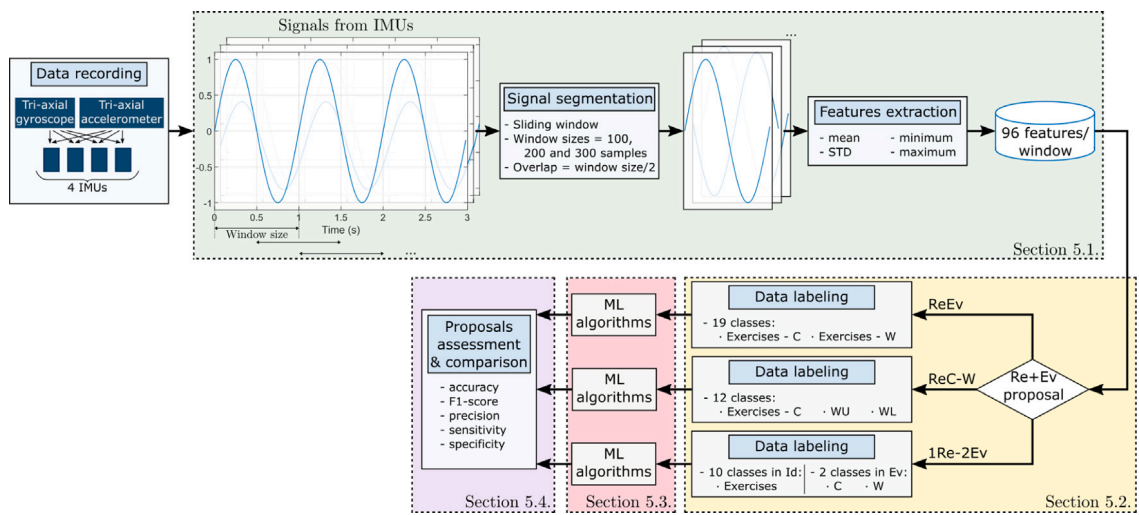

**Fig. 4.** Flowchart of the data analysis. Using the IMU signals, we split them in windows and extract, from each one, its mean, average, minimum and maximum. Then, we label the data according to the corresponding proposal. The labeled data are the inputs in the ML algorithms, which are then assessed using five metrics. Each process box, squared with dashed lines, indicates the corresponding section of this document in which this process is detailed.

*5.1. Signal processing*

In this study, we use a sliding window to segment the raw IMU signals, as depicted in the green rectangle of Fig. 4. Three window sizes are analyzed: 100, 200 and 300 samples, which correspond to windows of 1, 2 and 3 s, respectively. We limit the window size to 300 in order to find a balance between the algorithms' performance and the possible motions included in one single window. The larger the window, the more likely to mix different motions in them. Also, according to Banos et al. (2014), the interval 1–2 s was proved to provide the best trade-off between recognition speed and accuracy when a high variety of features is employed. We limit the interval to 3 s because we use simple features in our analysis and they proved a direct relation between the features and the window size. In all cases, the overlap between consecutive windows is set to the 50 % of the window size. We use four time-domain features commonly found in the literature: mean, standard deviation, maximum and minimum of signals over each axis. Since motions were recorded with four IMUs that record six signals each, the data set includes 96 features per window.

*5.2. Data labeling*

The proposals are based on supervised algorithms, that require the labeling of samples for their training. We manually selected the beginning and the end of each exercise series and we associated the corresponding label to each group of features. We know the exercise performed and its quality by the design of the experiments and their supervision, as explained in Section 4.3. However, the separation of the data between the different classes depends on the proposed method, since their output labels differ. It implies that the data labeling depends on the proposal and, more specifically, on the definitions of the classes into which the data are to be divided, as shown in the yellow rectangle of Fig. 4. In the following, we explain this labeling process together with the labels assigned to each data class.

*5.2.1. ReEv*

In this proposal, labels include information of the kind of exercise and whether the exercise is correctly or wrongly performed, e.g. the EAH exercise, divides into EAH-C and EAH-W, which correspond to its correct and wrong performances. The only exercise that does not include the performance information is GHT because its wrong performance is included in the wrong performance of normal gait, so it is considered as the wrong GAT exercise. We study eight exercises, and two of them are divided into the right or left sides of performance. As ReEv separately considers their correct and wrong performances, this proposal includes 19 classes, 9 kinds of exercises with their correct and wrong performance, and GHT.

*5.2.2. ReC-W*

The labels of the second proposal for the properly performed motions include the information about which exercise is being carried out. On the contrary, the wrong performances of the exercises are considered as only two kind of motions, those performed with the upper-limbs, labeled as WU, and those performed with the lower-limbs, labeled as WL. In this way, this proposal contains 12 classes (10 kinds of exercises, WU and WL).

Another main difference with the other two proposals is that in the design of ReC-W, we establish the number of samples in WU and WL (it means that we do not use the total number of samples, but we establish a quantity of them in order to ensure balance and variability). The other two methods include in their wrong-labeled classes those number of samples in which each exercise is wrongly performed because it is already balanced with the correct-labeled classes. In ReC-W, the number of samples in WU and WL has to ensure that the training set of the ML algorithms includes a high variability of wrongly performed motions so the test set can be properly classified. To do so, we separately double the number of samples in the largest class of the upper-and lower-limb exercises to establish the number of samples in WU and WL.

*5.2.3. 1Re-2Ev*

Since this proposal consists of two classification stages, we separate the labels depending on the objective of each stage. In the recognition stage, labels correspond to the kind of exercise, so we have 10 classes, whereas in the evaluation stage, the two possible labels indicate only if the classification is correctly (C) or wrongly (W) performed.

*5.3. Classifier training and validation*

After the extraction of the features and the labeling of associated windows, we trained the algorithms detailed in Section 3.3. For the optimization of hyperparameters, we split data into training set and validation set. The training set includes the data of twenty-four volunteers, the validation set includes the data of five volunteers. In this way, we have one volunteer left in order to test the algorithms through a LOSO cross-validation (Arlot & Celisse, 2010), as explained in the following section. We study these algorithms using the Matlab R2020b software as follows:

- We evaluate SVMs with three different kernel mappings: linear ($SVM_L$), polynomial ($SVM_P$) and Gaussian ($SVM_G$).
- We optimize the minimum leaf size of DT.
- We validate the number of neighbors in KNN, analyzing from 1 to 20 of them and selecting the one that obtains the highest accuracy in the validation test.
- In MLP, we set the number of hidden layers to 1 since we do not observe any improvement of performance using more than one. We set the number of neurons as the average of the number of inputs (which is always 96 features), being the outputs the number of classes identified by the proposals, according to Mukhopadhyay (2018).
- Since ELM is a fast algorithm, we validate the number of neurons from 10 to 1000 neurons and select the number that reports the highest accuracy.

### 5.4. Proposals assessment

The proposals are evaluated with a LOSO cross-validation, using the data of each volunteer as test data set, so the training-test process is carried out 30 times, one for each volunteer. This is the most robust type of cross-validation in studies that involve human subjects, because it allows the subject-to-subject variability and avoids the autocorrelation in time series data obtained with one subject. It is also a more demanding cross-validation method than k-fold or random cross validation, so its results are expected to be worse than with those last types of cross-validation.

The assessment of the proposals is carried out in terms of the average metrics for the thirty volunteers. This is the final step, highlighted with a purple rectangle in Fig. 4. The metrics considered operate separately for each class, which correspond to the kind of exercise and its performance quality. In this way, the positives ($P$) of a class are their number of sample, and the negatives ($N$) are the samples that correspond to the rest of classes. Positives are divided into true positives ($TP$) and false positives ($FP$) according to whether the samples really belong to the recognized class or they are misclassified, respectively. Negatives are also divided intro true negatives ($TN$) and false negatives ($FN$). $TN$ are those samples which do not belong to the considered class whereas $FN$ refer to the samples which really are members of the considered class but they are wrongly classified.

We study the proposals in terms of their accuracy (2), which measure the percentage of cases that the model has correctly predicted, although it must be combined with other different metrics for a meaningful analysis of the model. We also use the precision (3) to measure the quality in the detection of each class; sensitivity (4) to determine the effectiveness in the identification of each class; the F1-score (5) that combines those two previous metrics assuming that both of them are equally important; and finally the specificity (6) relates to the model's ability to correctly classify a sample that does not correspond to a class in this way, so it measures the ability to detect negative labels.

$$\text{acc}(\%) = \frac{TP+TN}{P+N} 100 \quad (2)$$

$$\text{prec}(\%) = \frac{TP}{TP+FP} 100 \quad (3)$$

$$\text{sens}(\%) = \frac{TP}{TP+FN} 100 \quad (4)$$

$$\text{F1}(\%) = \frac{2TP}{2TP+FP+FN} 100 \quad (5)$$

$$\text{spec}(\%) = \frac{TN}{TN+FP} 100 \quad (6)$$

We study the average values for all classes and volunteers. In order to provide an in-depth study on the results, including the differences in the identification quality of each class, we also analyze the confusion matrices.

## 6. Experimental results

In this section, we describe and discuss the results of the proposals for the recognition and evaluation of exercises. We initially analyze the best window lengths and ML algorithms for the proposals in Section 6.1. Then, we separately provide a detailed study of the three proposals. Section 6.2 contains the results obtained using ReEv, combining both classifications in one single step. Section 6.3 includes the results of ReC-W, which recognizes the correctly performed exercises and labels the wrong performances as WU or WL, depending on whether the motions are performed with the upper-or lower-limbs. Section 6.4 details the results of 1Re-2Ev, the two-stage method that firstly recognizes the exercises and secondly evaluates them. Finally, Section 6.5 provides a comparison of the three proposals.

### 6.1. Windows length analysis and ML algorithms performance

In order to analyze the most suitable window length for the proposals and the best ML algorithm, we separately study the metrics with each proposal. For the case of ReEv, we provide in Table 1 its resultant accuracy, F1-score, precision, sensitivity and specificity, using the three window sizes of 100, 200 and 300 samples. The first column for each window size shows the accuracy of each method, which in most of cases is about 80 %. As expected, the method that obtains the lowest metrics is DT, which is used as baseline, resulting in an accuracy lower than the 80 % with all the window sizes. This poor results are mainly caused by its common overfitting problem. Conversely, the two best algorithms are $SVM_L$ and RF. We focus on $SVM_L$ to study the best window size because it gives the highest metrics in most cases. All its metrics improve with the enlargement of the window size from 100 to 300 samples. The accuracy increases from 83.5 % to 88.3 %, the F1-score from 84.0 % to 88.1 %, the precision from 83.9 % to 88.6 % and the sensitivity from 83.9 % to 89.2 %. So the most suitable window size is 300 samples. Conversely, specificity is above 99 % with all the ML methods and windows sizes, so this metric cannot be the criteria to choose the best proposal. This specificity value is due to the fact that the number of false positives of a single class is highly lower than its number of true negatives, which are the addition of the other correctly labeled exercises.

For the case of ReC-W, Table 2 shows its metrics. The best algorithms are $SVM_L$ and RF, whose results are written in green bold, and the worst is DT, whose results are highlighted in red. As in the previous method, the algorithm that obtains the best metrics is $SVM_L$, which range between 87.2 % and 99.2 %, improving with the window lengthening. Using $SVM_L$ the best metrics are obtained with a window size of 300 samples, it reaches an accuracy of 91.4 %, being the rest metrics about 90.7 %.

Finally, we analyze the results of 1Re-2Ev. Since this method divides the recognition and the evaluation of exercises into two different classifications, we study their results separately. Firstly, we evaluate the initial stage, the exercise recognition, in which each class collects the correct and wrong performances of the corresponding exercises. Secondly, we focus on the last stage, the exercise evaluation, whose inputs are the recognized exercises of the previous stage and classifies these exercises between correct or wrong.

Table 3 shows the resultant metrics of the recognition stage. The two best algorithms, printed its metrics in bold green, are SVM with the Gaussian and the polynomial kernels. These configurations obtain results above 95.2 % with all the window sizes. The kernel difference with respect to the previous methods is remarkable. It means that gathering the correct and wrong performance of each exercise into one class, what implies an increment of variability in each class, the data distribution in the classes changes and their best separation is no longer with a linear hyperplane. The increase in variability also makes more difficult the generalization of the exercise recognition with RF, whose metrics are no longer included between the best ones. The worst algorithm is DT again, written in red in Table 3, as expected because of its results in the previous proposals.

**Table 1**
Classification results obtained with ReEv expressed in terms of accuracy, F1-score, precision, sensitivity and specificity. The first row specifies the window size for the signals cutting to obtain those metrics. The highest metrics of the two best methods are in bold and green, and the lowest metrics are in red color.

|  | Window size = 100 samples | | | | | Window size = 200 samples | | | | | Window size = 300 samples | | | | |
| --- | --- | --- | --- | --- | --- | --- | --- | --- | --- | --- | --- | --- | --- | --- | --- |
|  | acc (%) | F1 (%) | prec (%) | sens (%) | spec (%) | acc (%) | F1 (%) | prec (%) | sens (%) | spec (%) | acc (%) | F1 (%) | prec (%) | sens (%) | spec (%) |
| $SVM_G$ | 82.0 | 82.0 | 81.8 | 82.6 | 99.0 | 85.8 | 86.2 | 85.3 | 86.2 | 99.2 | 87.4 | 87.9 | 86.8 | 88.4 | 99.3 |
| $SVM_L$ | 83.5 | 84.0 | 83.9 | 83.9 | 99.1 | 87.4 | 88.1 | 87.7 | 88.6 | 99.3 | 88.3 | 89.8 | 88.6 | 89.2 | 99.4 |
| $SVM_P$ | 82.7 | 82.6 | 82.6 | 83.0 | 99.0 | 86.1 | 86.9 | 86.3 | 87.6 | 99.2 | 87.5 | 88.3 | 87.6 | 88.7 | 99.3 |
| RF | 83.6 | 83.1 | 83.7 | 83.6 | 99.1 | 86.3 | 86.5 | 86.7 | 87.0 | 99.2 | 88.4 | 88.8 | 88.7 | 89.4 | 99.4 |
| KNN | 77.1 | 77.2 | 77.5 | 76.6 | 98.7 | 80.6 | 81.3 | 81.4 | 80.8 | 98.9 | 84.1 | 85.1 | 84.9 | 84.8 | 99.1 |
| ELM | 82.1 | 81.5 | 82.2 | 80.4 | 99.0 | 85.0 | 84.5 | 85.2 | 85.0 | 99.2 | 88.0 | 88.2 | 87.8 | 88.5 | 99.3 |
| MLP | 81.2 | 81.1 | 81.7 | 80.1 | 99.0 | 81.7 | 83.0 | 81.0 | 82.7 | 99.0 | 81.9 | 84.2 | 81.6 | 83.5 | 99.0 |
| DT | 69.3 | 69.7 | 70.0 | 69.2 | 98.3 | 72.8 | 72.9 | 73.7 | 73.9 | 98.5 | 75.5 | 75.5 | 76.2 | 76.3 | 98.6 |

**Table 2**
Classification results obtained with ReC-W expressed in terms of accuracy, F1-score, precision, sensitivity and specificity. The first row specifies the window size for the signals cutting to obtain those metrics. The highest metrics of the two best methods are in bold and green, and the lowest metrics are in red color.

|  | Window size = 100 samples | | | | | Window size = 200 samples | | | | | Window size = 300 samples | | | | |
| --- | --- | --- | --- | --- | --- | --- | --- | --- | --- | --- | --- | --- | --- | --- | --- |
|  | acc (%) | F1 (%) | prec (%) | sens (%) | spec (%) | acc (%) | F1 (%) | prec (%) | sens (%) | spec (%) | acc (%) | F1 (%) | prec (%) | sens (%) | spec (%) |
| $SVM_G$ | 86.3 | 85.6 | 84.4 | 90.4 | 98.7 | 86.7 | 88.1 | 84.5 | 91.0 | 98.8 | 88.4 | 89.9 | 84.8 | 92.1 | 99.0 |
| $SVM_L$ | 87.2 | 88.2 | 88.5 | 89.1 | 98.8 | 89.9 | 90.9 | 90.0 | 91.7 | 99.1 | 91.4 | 92.6 | 90.7 | 92.9 | 99.2 |
| $SVM_P$ | 85.8 | 85.9 | 84.8 | 88.6 | 98.7 | 88.1 | 89.6 | 87.5 | 91.6 | 98.9 | 89.3 | 89.9 | 87.4 | 92.4 | 99.1 |
| RF | 87.0 | 87.3 | 87.5 | 89.5 | 98.8 | 89.5 | 90.0 | 89.6 | 92.1 | 99.0 | 89.4 | 89.8 | 88.7 | 92.3 | 99.1 |
| KNN | 80.6 | 82.3 | 83.4 | 82.4 | 98.2 | 84.5 | 86.7 | 87.1 | 85.7 | 98.5 | 85.8 | 87.5 | 88.0 | 86.5 | 98.7 |
| ELM | 86.0 | 87.7 | 89.8 | 85.7 | 98.7 | 88.8 | 90.4 | 91.6 | 89.5 | 98.9 | 89.4 | 90.3 | 91.5 | 90.2 | 99.0 |
| MLP | 85.4 | 85.9 | 86.9 | 85.7 | 98.6 | 85.0 | 87.7 | 86.9 | 87.0 | 98.6 | 85.1 | 88.7 | 86.7 | 87.1 | 98.7 |
| DT | 75.7 | 77.0 | 76.4 | 79.5 | 97.7 | 79.1 | 80.1 | 79.6 | 82.1 | 98.0 | 77.2 | 77.6 | 76.6 | 80.5 | 98.0 |

**Table 3**
Classification results obtained in the first stage of 1Re-2Ev expressed in terms of accuracy, F1-score, precision, sensitivity and specificity. The first row specifies the window size for the signals cutting to obtain those metrics. The highest metrics of the two best methods are in bold green and the lowest metrics are in red.

|  | Window size = 100 samples | | | | | Window size = 200 samples | | | | | Window size = 300 samples | | | | |
| --- | --- | --- | --- | --- | --- | --- | --- | --- | --- | --- | --- | --- | --- | --- | --- |
|  | acc (%) | F1 (%) | prec (%) | sens (%) | spec (%) | acc (%) | F1 (%) | prec (%) | sens (%) | spec (%) | acc (%) | F1 (%) | prec (%) | sens (%) | spec (%) |
| $SVM_G$ | 95.4 | 95.6 | 95.1 | 92.1 | 99.5 | 95.9 | 96.1 | 95.6 | 96.2 | 99.5 | 96.2 | 96.4 | 96.0 | 95.7 | 99.6 |
| $SVM_L$ | 94.8 | 94.8 | 94.7 | 92.0 | 99.4 | 95.8 | 95.8 | 95.7 | 95.5 | 99.5 | 96.0 | 96.0 | 95.9 | 95.8 | 99.6 |
| $SVM_P$ | 95.2 | 95.1 | 95.0 | 92.6 | 99.5 | 95.8 | 95.8 | 95.7 | 95.0 | 99.5 | 96.1 | 96.1 | 96.0 | 94.8 | 99.6 |
| RF | 93.8 | 93.8 | 93.6 | 91.5 | 99.3 | 94.6 | 94.6 | 94.4 | 93.3 | 99.4 | 95.1 | 95.2 | 94.8 | 94.4 | 99.5 |
| KNN | 92.0 | 92.2 | 92.2 | 88.5 | 99.1 | 92.8 | 92.9 | 93.0 | 90.9 | 99.2 | 94.4 | 94.9 | 94.7 | 92.7 | 99.4 |
| ELM | 93.4 | 93.2 | 93.0 | 90.3 | 99.3 | 95.4 | 95.5 | 95.2 | 93.8 | 99.5 | 95.8 | 95.6 | 95.5 | 95.1 | 99.5 |
| MLP | 94.3 | 94.4 | 94.3 | 90.5 | 99.4 | 94.9 | 94.9 | 94.8 | 93.9 | 99.4 | 95.5 | 95.7 | 95.3 | 94.9 | 99.5 |
| DT | 87.5 | 87.9 | 87.7 | 84.4 | 98.6 | 89.8 | 89.9 | 89.8 | 87.4 | 98.8 | 89.9 | 89.8 | 89.9 | 87.9 | 98.9 |

With regard to the windows length, even though 1Re-2Ev provides a high accuracy with all the window sizes, using windows of 300 samples only improves 1 % the results obtained with the smallest size. However, this method allows us to determine the type of exercise being performed with good metrics by using the smallest window (100 samples). In this way, the most appropriate window size of 1Re-2Ev will be given either by the system requirements or by the second stage of the method.

In the exercises evaluation, carried out after their recognition, we focus on the accuracy as main metric and on the F1-score since it combines the precision and sensitivity. Table 4 shows these metrics for each of the evaluated exercises with the considered ML algorithms.

Contrary to the results in the exercises recognition, in the evaluation stage, the metrics noticeably improve with the window lengthening. With the lowest windows, most algorithms achieve an accuracy between 90 % and 95 %, whereas with the largest ones, most algorithms show an accuracy above 95 %, reaching even some perfect classifications. Because of these results, the optimal window size for 1Re-2Ev is 300 samples, despite of the stage can be performed with smallest windows.

The worst algorithm with this approach differs from the other two proposed approaches. In this case, the ELM results in the poorest metrics, which are marked in red in Table 4. As in the case of DT, the main reason is its limited capacity of generalization.

The best results obtained with the 300-sample windows, marked in green in Table 4, are achieved with different variations of the SVM, whose results are shown in Fig. 5 in order to ease their interpretation. Specifically, the Gaussian kernel, the one presented in blue, is the most suitable for the three window sizes, overcoming the other methods, in most of cases. However, this two-stage approach allows us to use a different algorithm depending on the recognized exercise. It is interesting because $SVM_G$ is in most cases the best algorithm, as seen in Fig. 5, but there are three exercises in which another kernel overcomes its accuracy. It is the case of he GAT, HAR and SQT exercises, whose evaluation is better with a linear kernel than with a Gaussian one.

Although the larger the window size, the better results, it is convenient to analyze whether a longer length could result in a significant improvement of the metrics. To do so, we study the metrics improvement with each window size enlargement. Using ReEv with $SVM_L$, its metrics improve around 4 % with the first lengthening, changing the window size from 100 samples to 200 samples, see Table 1. With the second lengthening, from 200 samples to 300 samples, metrics improve only around 1 %. The same differences in the improvement of metrics can be found in the results of the other three methods, included in Tables 2–4. According to the methods' metrics, enlarging the window size from 100 samples to 200 samples, we obtain the highest improvement. Then, we can conclude that windows above 300 samples will not significantly improve these results. In addition, we can consider 3 s (i.e. 300 samples) as the window size limit to avoid that several movements occur in the same window or are performed differently. Although increasing the window may be beneficial in terms of the evaluated metrics, temporal resolution of the execution of the exercises can be lost.

**Table 4**
Classification results obtained in the second stage of 1Re-2Ev expressed in terms of accuracy, F1-score, precision, sensitivity and specificity. The first row specifies the window size for the signals cutting to obtain those metrics. The highest metrics of the two best methods are in bold green and the lowest metrics are in red.

| Window size = 100 samples | EAH | | EFE | | SQZ | | GAT | | HAL | | HAR | | KFL | | KFR | | SQT | |
|---|---|---|---|---|---|---|---|---|---|---|---|---|---|---|---|---|---|---|
| | acc (%) | F1 (%) | acc (%) | F1 (%) | acc (%) | F1 (%) | acc (%) | F1 (%) | acc (%) | F1 (%) | acc (%) | F1 (%) | acc (%) | F1 (%) | acc (%) | F1 (%) | acc (%) | F1 (%) |
| SVM G | **94.8** | **94.7** | **94.2** | **94.3** | **87.3** | **85.8** | **87.7** | 82.9 | **93.4** | **92.8** | **88.2** | **87.9** | **95.9** | **95.8** | 93.7 | 93.5 | 81.2 | 78.9 |
| SVM L | 90.8 | 90.6 | 87.9 | 87.6 | 82.8 | 80.5 | **87.2** | **83.1** | **95.6** | **95.4** | **88.9** | **88.7** | 95.8 | 95.6 | **97.9** | **97.7** | **84.9** | **83.4** |
| SVM P | 91.7 | 91.7 | 93.6 | 93.5 | 83.8 | 83.3 | 86.3 | 80.6 | 92.5 | 91.8 | 86.5 | 85.5 | 97.2 | 97.1 | 95.2 | 95.0 | 80.1 | 78.7 |
| RF | **94.0** | **93.9** | **94.0** | **93.7** | **84.7** | **82.4** | 85.8 | 81.9 | 94.0 | 94.2 | 85.5 | 84.1 | 95.9 | 95.7 | 94.0 | 92.4 | **81.6** | **79.7** |
| KNN | 87.5 | 87.4 | 89.0 | 88.5 | 78.2 | 77.0 | 78.7 | 73.3 | 88.4 | 87.7 | 82.6 | 81.8 | 86.4 | 85.8 | 84.3 | 83.4 | 75.2 | 73.0 |
| ELM | <span style="color:red">79.9</span> | <span style="color:red">79.4</span> | <span style="color:red">72.7</span> | <span style="color:red">73.7</span> | <span style="color:red">71.2</span> | <span style="color:red">69.0</span> | <span style="color:red">79.6</span> | <span style="color:red">74.5</span> | <span style="color:red">82.5</span> | <span style="color:red">83.2</span> | <span style="color:red">77.8</span> | <span style="color:red">76.4</span> | <span style="color:red">81.2</span> | <span style="color:red">80.3</span> | <span style="color:red">78.3</span> | <span style="color:red">77.2</span> | <span style="color:red">72.3</span> | <span style="color:red">69.7</span> |
| MLP | 90.0 | 89.6 | 88.1 | 87.9 | 83.7 | 82.7 | 85.7 | 82.0 | **93.4** | 92.6 | 87.2 | 85.8 | 95.7 | 95.6 | 92.2 | 91.6 | 82.7 | 81.5 |
| DT | 84.9 | 84.6 | 87.4 | 88.4 | 74.3 | 72.1 | 79.5 | 74.5 | 87.2 | 86.3 | 75.7 | 74.6 | 85.2 | 84.6 | 82.5 | 80.6 | 70.5 | 68.3 |
| Window size = 200 samples | EAH | | EFE | | SQZ | | GAT | | HAL | | HAR | | KFL | | KFR | | SQT | |
| | acc (%) | F1 (%) | acc (%) | F1 (%) | acc (%) | F1 (%) | acc (%) | F1 (%) | acc (%) | F1 (%) | acc (%) | F1 (%) | acc (%) | F1 (%) | acc (%) | F1 (%) | acc (%) | F1 (%) |
| SVM G | **98.4** | **98.4** | **97.8** | **97.7** | **90.6** | **89.5** | 89.7 | 85.8 | **97.5** | **97.4** | 92.4 | 91.8 | **100.0** | **100.0** | **98.1** | **98.1** | **92.2** | **91.8** |
| SVM L | 96.6 | 96.5 | 94.2 | 93.9 | 87.8 | 7.5 | **91.4** | **89.0** | 95.0 | 95.1 | **94.0** | **94.3** | **100.0** | **100.0** | **98.7** | **98.7** | 88.2 | 87.0 |
| SVM P | **98.4** | **98.4** | 95.6 | 95.5 | **90.9** | **90.2** | **90.4** | **87.4** | 96.7 | 96.5 | 93.1 | **93.0** | **100.0** | **100.0** | 95.6 | 95.5 | **90.9** | **90.2** |
| RF | **98.7** | **98.7** | **96.0** | **95.9** | 87.1 | 86.5 | 90.0 | 86.9 | 95.9 | 95.8 | **93.6** | 92.3 | 99.1 | 99.0 | 95.5 | 94.7 | 86.8 | 86.7 |
| KNN | 93.0 | 92.8 | 94.6 | 94.3 | 83.4 | 83.4 | 82.7 | 78.3 | 95.3 | 95.1 | 90.6 | 89.1 | 94.8 | 94.6 | 88.6 | 88.2 | 79.6 | 78.5 |
| ELM | <span style="color:red">84.2</span> | <span style="color:red">84.1</span> | <span style="color:red">79.6</span> | <span style="color:red">78.6</span> | <span style="color:red">71.0</span> | <span style="color:red">71.4</span> | <span style="color:red">84.8</span> | <span style="color:red">80.4</span> | <span style="color:red">88.8</span> | <span style="color:red">88.6</span> | <span style="color:red">79.4</span> | <span style="color:red">80.3</span> | <span style="color:red">87.7</span> | <span style="color:red">87.5</span> | <span style="color:red">80.2</span> | <span style="color:red">78.7</span> | <span style="color:red">76.1</span> | <span style="color:red">74.7</span> |
| MLP | 95.8 | 95.7 | 93.2 | 93.4 | 83.9 | 82.6 | 87.6 | 84.1 | **98.7** | **98.6** | 89.8 | 88.5 | 98.8 | 98.7 | 95.5 | 95.3 | 85.8 | 84.2 |
| DT | 90.0 | 89.8 | 93.2 | 93.5 | 80.6 | 79.1 | 76.6 | 73.0 | 83.7 | 84.2 | 83.9 | 83.7 | 93.4 | 93.1 | 82.7 | 82.2 | 72.1 | 69.6 |
| Window size = 300 samples | EAH | | EFE | | SQZ | | GAT | | HAL | | HAR | | KFL | | KFR | | SQT | |
| | acc (%) | F1 (%) | acc (%) | F1 (%) | acc (%) | F1 (%) | acc (%) | F1 (%) | acc (%) | F1 (%) | acc (%) | F1 (%) | acc (%) | F1 (%) | acc (%) | F1 (%) | acc (%) | F1 (%) |
| SVMG | **100.0** | **100.0** | **100.0** | **100.0** | **96.3** | **96.0** | 91.3 | 88.0 | **100.0** | **100.0** | 92.4 | 91.9 | **100.0** | **100.0** | 99.3 | 99.3 | **95.1** | **94.8** |
| SVM L | **100.0** | **100.0** | 99.1 | 99.0 | 90.1 | 88.8 | **92.5** | **90.5** | 96.8 | 96.6 | **95.6** | **95.5** | **100.0** | **100.0** | **100.0** | **100.0** | **97.0** | **97.3** |
| SVM P | **100.0** | **100.0** | **100.0** | **100.0** | 92.4 | 92.0 | **93.6** | **92.2** | 99.0 | 99.0 | 92.9 | 91.8 | **100.0** | **100.0** | **99.8** | **99.8** | 94.0 | 93.5 |
| RF | **100.0** | **100.0** | **100.0** | **100.0** | **94.4** | **94.0** | 89.0 | 86.3 | 98.8 | 98.8 | **93.5** | **92.8** | 99.3 | 99.3 | 98.2 | 98.4 | 92.1 | 91.5 |
| KNN | 96.9 | 96.8 | 97.6 | 97.4 | 82.9 | 82.9 | 84.4 | 80.4 | 95.1 | 95.2 | 82.4 | 83.4 | 98.4 | 98.3 | 96.7 | 96.6 | 87.9 | 87.2 |
| ELM | <span style="color:red">86.1</span> | <span style="color:red">86.0</span> | <span style="color:red">81.0</span> | <span style="color:red">80.5</span> | <span style="color:red">73.9</span> | <span style="color:red">74.6</span> | <span style="color:red">85.2</span> | <span style="color:red">81.5</span> | <span style="color:red">85.0</span> | <span style="color:red">84.3</span> | <span style="color:red">78.5</span> | <span style="color:red">78.4</span> | <span style="color:red">86.9</span> | <span style="color:red">86.2</span> | <span style="color:red">78.2</span> | <span style="color:red">77.1</span> | <span style="color:red">76.6</span> | <span style="color:red">74.9</span> |
| MLP | **100.0** | **100.0** | 98.1 | 98.0 | 86.9 | 87.2 | 90.3 | 86.7 | **100.0** | **100.0** | 90.9 | 89.9 | 98.3 | 98.2 | 99.4 | 99.3 | 93.0 | 92.6 |
| DT | 93.9 | 93.7 | 88.1 | 87.6 | 84.9 | 83.1 | 77.1 | 73.5 | 87.5 | 87.0 | 84.8 | 83.1 | 95.0 | 94.8 | 86.5 | 86.6 | 75.9 | 73.7 |

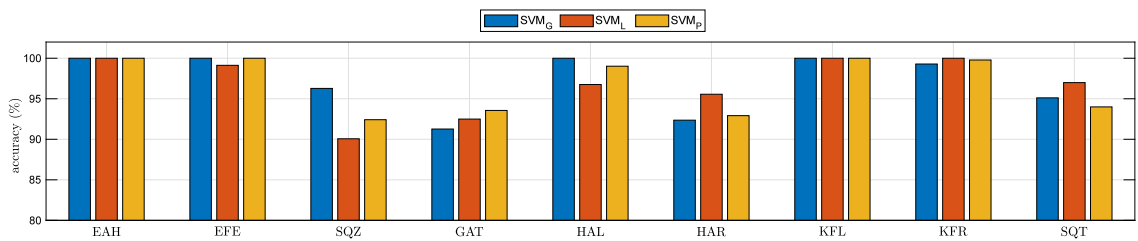

**Fig. 5.** Accuracy of the different variations of SVM with the 300-sample windows, separated by the exercises. The interval zooms in the interesting area in order to ease the comparison of the different algorithms performances. We depict SVM$_G$ in blue, SVM$_L$ in red and SVM$_P$ in yellow.

In Preatoni et al. (2020), they propose the use of 6 s windows, obtaining a similar accuracy as this work. Notice that the size of the window is the half. In this way, our signal processing includes less data so it is simpler. Conversely, we use an overlap of 50 % instead of 10 %, so we analyze each 1.5 s, whereas they provide information each second.

### 6.2. ReEv: Recognition and evaluation in a single step

The best configuration of ReEv, using SVM$_L$ and windows of 300 samples, gives an accuracy, F1-score, precision and sensitivity between 88.3 % and 89.8 %, see Table 1. These metrics imply that we obtain adequate results considering that we simplify the exercise characterization procedure to a single classification of nineteen classes with high variability.

In order to analyze in-depth the origin of errors and which are eliminated with the window lengthening, we study the average confusion matrix of the method that reports the best metrics, SVM$_L$. Fig. 6 shows these average confusion matrices using a window size of 100 and 300 samples, respectively. The first quadrant of both matrices includes the lowest amount of errors, so among the correct exercises properly evaluated, the SVM$_L$ rarely misrecognizes the exercises. The main error in this recognition is between the two kinds of gait, GAT, the normal and GHT, the heel-toe gaits, as marked with the green square in Fig. 6. These errors are a consequence of that GAT and GHT can seem similar exercises during some intervals of the motion.

The second and third quadrants with a window size of 100 samples contain the highest number of errors. Their distribution is specially remarkable since they form an almost diagonal line with the cells of misclassifications that correspond to each kind of exercise labeled as wrong whereas it is correct and the opposite. Again, the only exceptions are the highly related kinds of gait, GAT and GHT, which are marked with two green rectangles in Fig. 6. In this way, errors by ReEv do not combine misrecognitions of exercises with their incorrect evaluations, but they can be divided into these two sources of error. Most of the errors of these quadrants are mainly caused by the incorrect evaluations of the exercises as correct or wrong. These errors are still present when the window lengthens to 300 samples, however, they are diminished and some of them are even eliminated. Thus, the evaluation of exercises requires a higher window size than the recognition of the correctly performed exercises.

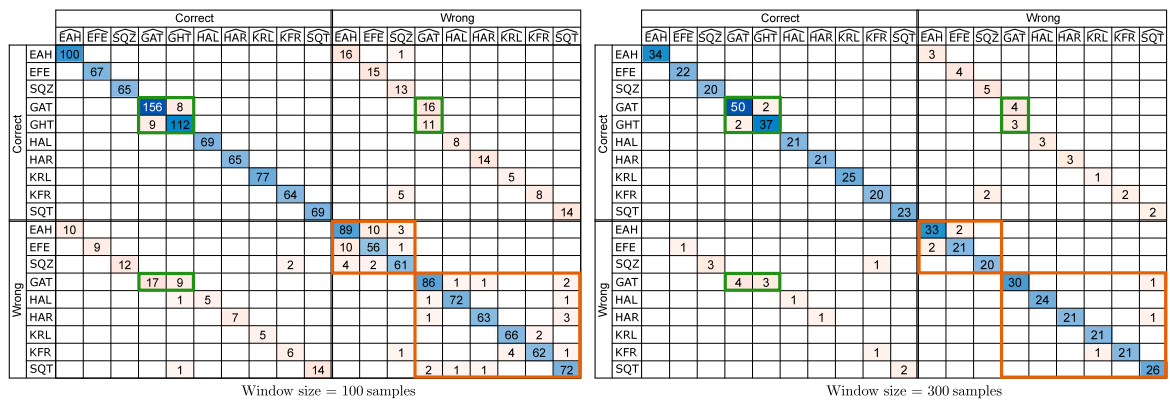

**Fig. 6.** Average confusion matrix for the 30 volunteers using ReEv with $SVM_L$ and a window size of 100 and 300 samples (left and right respectively). Columns include the predicted classes, which are pointed out with a ˆ over their labels; and rows contain the actual classes. Double lines divide the correctly and wrongly performed exercises. The blue highlighted cells correspond to the average number of correct classifications and the cream colored ones indicate the misclassifications.

It is remarkable that the errors from the evaluation as wrong of correctly performed exercises (32 misclassifications) double the errors from the evaluation as correct of wrongly performed exercises (17 misclassifications). It means that classifiers detect as wrong performance those exercises that due to the variability of volunteers some correctly exercises are labeled as wrong. The main reason of this error is the high amount of classes considered in this classification with respect to the amount of data, so each class includes so little variability that slight variations of the prescribed motions are classified as wrong performances.

The fourth quadrant using window sizes of 100 samples also has a great amount of errors. As hypothesized, wrongly performed exercises with the upper-or lower-limbs can be similar and, as consequence, they are misidentified. The errors distribution shows that the mis-recognitions are more common in the upper-limb exercises than in the lower-limb ones. This difference occurs because during the execution of upper-limb exercises, both arms can freely move in the 3D-space whereas in the lower-limb exercises the volunteers posture does not allow as much freedom of movement. For example, if they are seated for the KFL exercise, the posture is different if they are walking, so the accelerometer measurements are clearly different and even if both exercises are wrongly performed, their measurements differ. This change in the posture is also the reason why the methods do not mix up the wrong upper-and lower-limb exercises. In this way, the errors in the recognition of wrong exercises can be divided according to if they are lower-or upper-limb exercises, which are the orange squared cells in Fig. 6, fourth quadrant. These errors decreases using windows with 300 samples instead of 100 samples, but they maintain a similar distribution. Increasing the window size, wrongly performed exercises seem to be better identified. However, this reduction is mainly caused by the decrease of samples when increasing the window size, so a solution to this source of error is still needed. Thus, these results justify the next method, ReC-W, that groups the upper-and lower-limb wrongly performed exercises.

*6.3. ReC-W: Recognition of correct exercises and detection of the wrong ones*

ReC-W overcomes the limitation of ReEv in the recognition of wrong exercises by reducing the number of classes that include wrong performance of motions to only two: WU and WL. In this way, using $SVM_L$ and a window size of 300 samples, ReC-W reaches accuracy, F1-score, precision and sensitivity metrics between 90.7 % and 92.9 %, see Table 2. According to these metrics, ReC-W proves to be a competitive method for the combination of the exercises recognition and evaluation tasks although it loses the information related to which exercise has been wrongly performed.

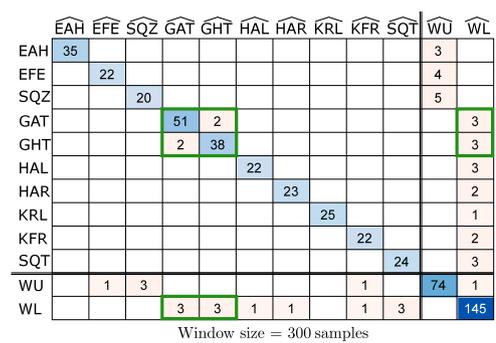

**Fig. 7.** Average confusion matrix for the 30 volunteers using ReC-W with $SVM_L$ and a window size of 300 samples. Columns divide the predicted classes, which are pointed out with a ˆ over their labels; and rows divide the actual classes. Double lines divide the correctly and wrongly performed exercises.

Fig. 7 depicts the confusion matrix obtained with the best method and configuration: $SVM_L$ and a window size of 300 samples. This confusion matrix shows that the recognition of correctly performed exercises is almost perfect, remaining the previous errors in gait types recognition, whose classification is marked in green. The division between upper-and lower-limb wrongly performed exercises, shown in the fourth quadrant of Fig. 7, has only one misrecognition out of 220 samples.

In this way, by gathering the wrong performed exercises in WU and WL, the only source of error still present in ReC-W is the one from the performance evaluation. That is the main reason why the metrics are improved with respect to the previous proposal, ReEv, as we will discuss in-depth in Section 6.5.

ReC-W is a promising approach with good metrics in the detection and evaluation of exercises, but it comes at a cost. We cannot obtain information about how the different wrong exercises are performed using ReC-W because all of them are gathered. As a consequence, the potential feedback of a combination of recognition and evaluation would be simpler than if we also knew which exercise is being performed wrongly. However, for applications aimed to count and characterize only the correct performances, this method has proven to be suitable.

*6.4. 1Re-2Ev: Recognition of exercises, followed by their evaluation*

We individually evaluate the results of each stage of this method: first, the ones corresponding to the exercises recognition and then,

**Fig. 8.** Average confusion matrix in the exercise recognition stage of 1Re-2Ev with SVM$_G$ and a window size of 300 samples. Columns divide the predicted classes, which are pointed out with a ˆ over their labels; and rows divide the actual classes.

those results obtained during the exercises evaluation. With regard to its metrics, with the exception of DT, most algorithms achieve accuracy, F1-score, precision, sensitivity and specificity above 90 % even with the smallest window, see Table 3. It implies that the recognition of the exercises, even combining their correct and wrong performance is highly accurate. The main reason is that 1Re-2Ev considers a lower number of classes with a high number of samples per class than the previous methods, so these data have a high variability in the training data. In this way, the distribution of the test data set, that belongs to a completely new volunteer, is more likely to be similar to the training data than when using less data per class.

We use a window size of 300 samples with the SVM$_G$ to obtain the confusion matrix shown in Fig. 8. As in the results of the previous proposals, most errors of 1Re-2Ev are located in the upper-limb exercises and in gait variations. It is consistent with the previous statements about the influence of the posture in the recognition results.

The results of this recognition stage, shown in Table 3, are comparable to the state-of-the-art methods for exercises recognition. For example, Zhao and Chen obtained an average accuracy of 96 % in the recognition of four basketball motions using four IMUs on the upper-limbs (Zhao & Chen, 2020). As in that reference, the best accuracy is obtained by using SVMs. Even if these motions are more complex than the one studied in this work, they use a four-fold cross-validation method, which is expected to give higher metrics than using a LOSO cross-validation, as the one used in this study. Similar accuracy is reported in Preatoni et al. (2020), where four fitness exercises are recognized within a continuous workout using five IMUs placed on the lower back, the upper-and lower-limbs of one side of the body. They report an accuracy between 94 % and 99 %, a precision between 89 % and 94 %, and a sensitivity between 79 % and 97 %, which are specially interesting because they recognize the transition intervals when no exercise is being performed. However, the fourteen participants in that study correctly performed all the evaluated motions. In our work we consider a higher motion variety. Finally, in Bavan et al. (2019), they use only one IMU placed on the arm for the motion monitoring. They obtain an accuracy about the 90 % using a ten-fold cross-validation that decreases to a maximum of 80 % when they use a LOSO cross-validation and RF. In this way, our exercise recognition obtains better metrics, mainly because we use four IMUs instead of one. We obtain competitive metrics in relation to the results in the literature, with an average about 96 % in the exercises recognition. In addition, we study and recognize a higher number of exercises, which include their correct and wrong performances.

For the exercises evaluation after their recognition, we initially analyze only the results of SVM$_G$. This algorithm provides an average accuracy and F1-score of 97.17 % and 96.67 %, respectively, see Table 4. Furthermore, adapting the most suitable ML algorithm for each exercise, the average accuracy and F1-score increase around 1 %, being 98.06 % and 97.89 %, respectively. It implies that the exercise evaluation obtains excellent metrics, close to perfect classifications, even including the initial error in the exercise recognition.

The results of the second stage, shown in Table 4, are comparable to the ones reported in the literature about exercises evaluation. The 90 % accuracy, using RF for the lunge evaluation in Whelan et al. (2016) and the 89 % accuracy using SVM for the single-leg squats exercises evaluation in Kianifar et al. (2017) are consistent with the results obtained in this work in the squat exercise, which is the most similar motion evaluated. For this exercise, we obtain slightly better accuracy, around 95 % with all SVMs configurations and 92 % with RF.

In the studies focused on the individual evaluation of multiple exercises, results are similar. The logistic regression in Giggins et al. (2014) achieved a maximum accuracy of 83 % on binary exercise evaluation of seven lower-limb exercises. In the exercises evaluation of Huang et al. (2016), the reported maximum accuracy was 97 %. These results are clearly comparable to the obtained in the lower-limb exercises evaluation in this work, which provide an accuracy above 95 %.

The results of the upper-limbs exercises are also comparable to those obtained in the literature. In Pereira et al. (2019), the fusion of two IMUs and sEMG sensors obtain an accuracy about 92 %. In this way, we obtain similar metrics in the exercises evaluation, although our classification is binary between correct and wrong, but we use less types of sensors. Similarly, in García de Villa et al. (2021), the authors found an accuracy between 98–99 % including both upper-and lower-limb exercises. These results are slightly better than the ones reported in the present work. However, the random cross-validation used in the previous work, is less demanding and the exercise evaluation is simpler when we have a previous knowledge about the exercise that is executed.

In order to compare the results of gait evaluation in Alcaraz et al. (2017) and our results, we consider that the wrongly performed gait is similar to an unhealthy gait. Therefore, our results, with an accuracy of 94 %, are in the same range of the results shown in Alcaraz et al. (2017). However, they obtain an accuracy of 100 % with LDA and NB by using more than four features to characterize the motions. Then, these results imply that the gait evaluation requires more features than the evaluation of other exercises to improve the obtained metrics.

In previous works (García de Villa et al., 2021), we also evaluated two kinds of gait, classifying as correct and wrong, with results around 98 %. However, in this work we obtain an accuracy of 93.6 % and an accuracy of 92.2 %. The increment of errors in the evaluation of gait can be derived from the including of GHT gait variation, which entails a higher variability of motions similar to gait. Also, in García de Villa et al. (2021) we used a random cross-validation with 10 iterations, whereas in this work we use LOSO cross-validation, so metrics are expected to decrease.

Finally, the differences of methods' performance between exercises are noteworthy. The evaluation of the simplest exercises achieves the best metrics. That is the case of EFE, EAH, HAL, KFL and KFR. The evaluation of these exercises produces an accuracy and a F1-score above 99 %. HAR is not included between them probably because of a bias generated by the order of the exercises in the experiments. During the firsts wrong repetitions of HAR, volunteers performed motions close to GAT or SQT, which were corrected by the time they did HAL. On the contrary, the evaluation of the most complex exercises, SQZ, GAT and SQT, has an accuracy and a F1-score between the 93 % and the 97 %. That is related to the easiness of separating between the correct and wrong performances of the simplest exercises, whose features clearly differ between a correct and a wrong performance. In the complex motions, features are more diverse and, as a consequence, they are closer in both performances than in simple motions.

*6.5. Comparison between the proposed methods*

One of the most remarkable similarities between the proposals is that SVM is the ML algorithm that provides the highest metrics in all of them, closely followed by RF. This algorithm is the most suitable one for the recognition and evaluation of the exercises studied in this work. The difference between the proposals is the kernel used. The linear one is the most appropriate for ReEv, ReC-W and the first stage of 1Re-2Ev, see Tables 1–3 whereas in the second stage of 1Re-2Ev the Gaussian and polynomial kernels provide better results, see Table 4.

Another interesting similarity is that the optimal window size is also common for all the proposals. The highest metrics are reported with the 3-second windows (300 samples).

Comparing the results of ReEv and ReC-W, shown in Tables 1 and 2 respectively, we can see that, as expected, ReC-W overcomes the initial proposal. Focusing on the best algorithm, $SMV_L$, with the largest window size, its accuracy, F1-score, precision and sensitivity increase a 3% with this change of approach in the classification, exceeding all of them the 90%. Conversely, the specificity remains almost similar but decreases with ReC-W for two reasons that can be seen in the confusion matrix of this method (see Fig. 7). Firstly, the number of $TN$ decreases for WU and WL, since they are the largest classes. WU and WL correspond to the last rows and columns in Fig. 7, so their true negatives are the first quadrant and the other corresponding class, i.e. WU for WL and the opposite. Secondly, their $FP$ increase because these errors correspond to the exercise evaluation, which correspond to the second and third quadrants in Fig. 7 and, as seen also in the results of ReEv, are the most frequent errors. In this way, errors in the recognition of wrongly performed exercises are eliminated using ReC-W. Not only the recognition of wrong exercises improves, but also the number of properly evaluated motions increases by decreasing the number of correct exercises labeled as wrong. We measure this improvement in terms of the F1-score of the WU and WL classes in ReC-W, compared to those classes of wrongly performed motions when using ReEv. Both the F1-scores of WU and WL are of 91% versus the average F1-score of the upper-and lower-limb related motions, which are 84% and 89%, respectively. So the increment of variability of motions in the WU and WL classes by using ReC-W decrease the evaluation errors, compared to the results obtained when identifying separately the wrongly performed exercises, as made with ReEv.

With regard to the comparison of ReEv and the first stage of 1Re-2Ev, we focus on the misrecognition errors shown in Fig. 6-right, resultant of ReEv using a window size of 300 samples, and the one in Fig. 8, that includes the results of the recognition stage of 1Re-2Ev. In Fig. 6-right, the misidentifications of ReEv are in the first and fourth quadrants and also in the second and third quadrants but only in these cells which do not belong to the main diagonal, e.g. when in the third quadrant GĤT is recognized instead of GAT. They sum a total of 20 misrecognitions by ReEv. Conversely, the first stage of 1Re-2Ev confuse 19 samples. That means a slight reduction of errors by the division of classifications caused by the aforementioned reduction of classes and increment of data variability in each of them.

These results in exercises recognition are not comparable to those of ReC-W because the recognition of ReC-W only includes the correctly performed motions so it is not the same input data and neither the same output information. ReC-W overcomes the errors of ReEv in the recognition of wrongly performed errors and improve its performance in the exercises evaluation. However, ReC-W only allows us to know if an exercise is wrongly performed, whereas ReEv and 1Re-2Ev give enough information to relate the wrong exercises with their characterization.

With respect to the exercise evaluation, in the second stage of 1Re-2Ev, most metrics are above 95% (see Table 4), which implies that 1Re-2Ev presents the lowest errors in the evaluation of exercises. This means that by separating both classifications, 1Re-2Ev overcomes the limitations related to the exercise evaluation of ReEv and ReC-W. This improvement in comparison with ReEv is caused by lowering the number of classes, whereas in comparison with ReC-W is a consequence of the reduction of variability in the wrong classes of each exercise.

In this way, 1Re-2Ev overcomes ReEv and ReC-W because of two main reasons: *(1)* the recognition metrics are better than ReEv and similar to ReC-W but gives more information since it also recognizes the wrong exercises and *(2)* it provides the highest metrics in the exercise evaluation. Also, this proposal includes the flexibility of tuning the algorithm for the exercise evaluation in its second stage, in order to optimize its results according to the recognized exercise.

One can argue that 4 features per signal and window (96 per window combining all sensors) are not enough to obtain the best performance of the ML algorithms. In fact, one of the alternatives to improve the obtained results is to increase the number of features. However, the features used in this work allow us to compare the proposals for the exercises recognition and evaluation, and to establish the most suitable approach for this complex task. In addition, we obtain high metrics with an accuracy about 91.4% with two of the proposed methods. These metrics prove that both proposals are comparable with the state-of-the-art methods even when they combine both tasks, recognition and evaluation of exercises, whereas in the literature these tasks are separately addressed.

Furthermore, the high metrics given by the three methods with such a variability of volunteers entail that the they adapt to different population. The main reason is that the design of motions is similar for all ages, so their correct performance is similar independently of the subject and they only show variations that are already in the analyzed database. In this way, the proposals are robust to changes in the motions caused by age.

**7. Conclusions**

This work proposes several approaches to automatically recognize and evaluate exercises included in a physical routine aimed for maintaining older people health status, what can prevent the onset of frailty. Our work contributes to the development of virtual coaches that help achieve healthy aging by supporting regular daily exercise, improving adherence to the physical routine and monitoring it. With the proposals, we demonstrate the feasibility of the characterization of this routine performance, which may become a reality in the near future.

For this complex task, we have proposed three alternatives: (1) identifying and evaluating in a single stage (ReEv); (2) identifying only the correct exercises (ReC-W) in a single stage, and (3) identifying in a first stage and then evaluating in a second stage whether the exercise is well or poorly performed (1Re-2Ev). These proposals have been evaluated in a set of 30 volunteers between 20 and 70 years old, with different ML algorithms. The metrics used to evaluate the proposals prove that the one-stage classification approaches are less suitable than the two-stage one. Combining the recognition and evaluation in a single classification problem, ReEv and ReC-W obtain an accuracy of 88% and 91%, depending on whether the classification of the wrong executions is performed. Conversely, the initial recognition followed by the exercises evaluation of 1Re-2Ev, gives an accuracy around the 95%, even with the error propagation from the first stage. 1Re-2Ev is also interesting since we prove that different exercises are evaluated better with different ML algorithm, and this approach allow us to assign the most suitable classifier to each performed exercise, after it has been recognized.

Another main difference between the one-and two-stage methods is that, even SVM is the most suitable algorithm in all the studied cases, the most suitable kernel differs between proposals. In the ReEv and ReC-W, $SVM_L$ overcomes the other methods, whereas in 1Re-2Ev, $SVM_G$ and $SVM_P$ provide the best results.

We also find that the recognition of correct motions is less demanding than their evaluation, i.e. in the evaluation, parameters as the window size are more relevant to obtain better results than in the exercises recognition. This is due to the differences in posture of the volunteers and, as a consequence, of the sensors, It helps to recognize the exercises but not to evaluate them. The most complex exercises, as GAT and SQZ, reported the worst metrics in their recognition and evaluation, so richer features should be used to improve their characterization.

As future work, we plan on reducing the number of sensors in order to evaluate a more user friendly sensory system, based on the best proposed method, 1Re-2Ev, and testing it with people older than the volunteers of this study. Finally, we will increase the features, specially in complex exercises, so that their metrics reach the values obtained with the simplest exercises evaluated.

### CRediT authorship contribution statement

**Sara García-de-Villa:** Conceptualization of this study, Methodology, Data curation, Software, Writing - Original draft preparation. **David Casillas-Pérez:** Conceptualization of this study, Writing - Original draft preparation. **Ana Jiménez-Martín:** Conceptualization of this study, Writing - Original draft preparation. **Juan Jesús García-Domínguez:** Conceptualization of this study, Writing - Original draft preparation.

### Declaration of competing interest

The authors declare that they have no known competing financial interests or personal relationships that could have appeared to influence the work reported in this paper.

### Data availability

Datasets related to this article can be found at https://zenodo.org/record/5052756#.YY5Xz2DMKUk, hosted at Zenodo.

### Acknowledgments

This work was supported by the Spanish Ministry of Science, Innovation and Universities (MICROCEBUS RTI2018-095168-B-C51) and the Youth Employment Program (PEJ-2020-PRE/TIC-17000).